\documentclass[a4paper,12pt]{article}
\begin{document}
\title{Simple analytic evaluation of the integrals in the Gaussian model of
Bose-Einstein correlations}
\author{K.Zalewski
\\ M.Smoluchowski Institute of Physics
\\ Jagellonian University, Cracow\footnote{Address: Reymonta 4, 30 059 Krakow,
Poland, e-mail: zalewski@th.if.uj.edu.pl}
\\ and\\ Institute of Nuclear Physics, Cracow}

\maketitle
\begin{abstract}
A generating function for products of Hermite polynomials is used to significantly
simplify the evaluation of the integrals $G_n(p,p')$ occurring in the Gaussian
model of multiple particle production. These integrals are crucial for studies of
multiparticle effects in Bose-Einstein correlations.
\end{abstract}

\noindent PACS numbers 25.75.Gz, 13.65.+i \\Bose-Einstein correlations, Gaussian
model. \vspace*{1cm}

 \textbf{1.} Bose-Einstein correlations in multiple particle production processes are
 a field of intense study. For recent reviews see \cite{BAY}-\cite{WEI}. An important
 subfield is the study of multiparticle  effects on multiplicity and momenta distributions.
 Here exactly soluble models are a convenient starting point. A soluble model,
 which is particularly popular, is the Gaussian model \cite{PRA} - \cite{WIE}. In
 this model all the integrations are Gaussian and consequently can be performed
 analytically. The generic integral is

\begin{equation}\label{}
  G_n(p,p') = \int \rho(p,k_1)dk_1\rho(k_2,k_3)\cdots dk_{n-1}\rho(k_{n-1},p'),
\end{equation}
where
\begin{equation}\label{matgau}
  \rho(p,p') = \frac{1}{\sqrt{2\pi\Delta^2}}\exp\left[-\frac{K^2}{2\Delta^2} -
  \frac{1}{2}R^2 q^2\right]
\end{equation}
with

\begin{equation}\label{parakq}
  K = \frac{p+p'}{2};\qquad q = p - p'.
\end{equation}
The constants $R^2$ and $\Delta^2$ are real and positive. For simplicity we start
with the one-dimensional case. In the Gaussian model the transition to three
dimensions is trivial. It is seen that the integrals $G_n$ are Gaussian and
consequently can be performed analytically. One needs, however, a closed formula
valid for any $n$. This is more difficult to obtain. Zim\'anyi and Cs\"org\"o
\cite{CZI,ZCS} following some suggestions from Pratt \cite{PRA} constructed a
system of recurrence relations for three series of parameters and after a rather
lengthy calculation got a formula for the integrals, which can be written in a
fairly simple form \cite{HSZ}. Wiedemann \cite{WIE} applied successfully the
MATHEMATICA program directly to the integrals $G_n$. In the present note we point
out that a very short derivation of the formula for the integrals $G_n$ can be
given, if a mathematical identity for Hermite polynomials is made use of. This
work is an extension of the method used in \cite{BZ1,BZ2}.

\textbf{2.} An identity, valid for $|z|< \frac{1}{2}$, is

\begin{equation}\label{idherm}
\Phi(x,y,z) \equiv \sum_{n=0}^\infty \frac{z^n}{n!}H_n(x)H_n(y) =
\frac{1}{\sqrt{1-4z^2}}\exp\left[ -4z\frac{z(x^2 + y^2) - xy}{1 - 4z^2}\right].
\end{equation}
A convenient starting point to check this formula is the definition of the Hermite
polynomials in the form \cite{RYG}\footnote{In this edition of the tables the
power of two is mistakenly written as $\frac{n}{2}$ instead of $n$. In order to
check the formula it is enough to expand the binomial into powers of $z$,
integrate term by term and compare the coefficients with the formulae given in
\cite{ABS}.}

\begin{equation}\label{}
  H_n(z) = \frac{2^n}{\sqrt{\pi}}\int_{-\infty}^{+\infty}(z + it)^n e^{-t^2}dt.
\end{equation}
After substituting it into (\ref{idherm}) the summation over $n$ gives an
exponential and it is enough to perform a double Gaussian integral to get the
right-hand side. The function $\Phi(x,y,z)$ is defined by this formula.

\textbf{3.} For any orthonormal set of functions $\psi_n(p)$ and any set of
nonnegative numbers $\lambda_n$ satisfying the normalization condition

\begin{equation}\label{}
  \sum_n \lambda_n = 1,
\end{equation}
the function

\begin{equation}\label{}
  \rho(p,p') = \sum_n\psi_n(p)\lambda_n \psi_n^*(p')
\end{equation}
can be interpreted as a well-defined single-particle density matrix in the
momentum representation. The functions $\psi_n$ are its eigenfunctions and the
numbers $\lambda_n$ are the corresponding eigenvalues. Let us choose in particular
for the functions $\psi_n$ the energy eigenfunctions in the momentum
representation of a one-dimensional harmonic oscillator and for the numbers
$\lambda_n$ a geometric progression \cite{BZ1,BZ2}:

\begin{equation}\label{}
  \psi_n(p) = \sqrt{\frac{\alpha}{\sqrt{\pi}2^n n!}}\exp\left[-\frac{\alpha^2
  p^2}{2}\right]H_n(\alpha p); \qquad \lambda_n = (1-z)z^n.
\end{equation}
It is assumed here that $|z| < 1$, so that the eigenvalues $\lambda_n$ are
correctly normalized. The parameter $\alpha$ is positive. Its interpretation in
terms of the mechanical parameters of the oscillator is of no importance here.
This choice corresponds to a single particle density matrix which using
(\ref{idherm}) can be written in the form

\begin{equation}\label{}
  \rho(p,p') = \frac{\alpha(1-z)}{\sqrt{\pi}}e^{-\frac{1}{2}\alpha^2(p^2 + p'^2)}\Phi(\alpha
  p, \alpha p', \frac{z}{2}),
\end{equation}
or substituting the definition of the function $\Phi$ and replacing the parameters
$p,p'$ by the parameters $K,q$ according to (\ref{parakq})

\begin{equation}\label{}
  \rho(K,q) = \frac{\alpha}{\sqrt{\pi}}\sqrt{\frac{1-z}{1+z}} \exp\left[ -\alpha^2
  K^2\frac{1-z}{1+z} - \frac{\alpha^2q^2}{4}\frac{1+z}{1-z}\right].
\end{equation}
This density matrix differs only by notation from the density matrix
(\ref{matgau}). The correspondence between the parameters is

\begin{equation}\label{}
  \alpha = \sqrt{\frac{R}{\Delta}};\qquad z = \frac{v-1}{v+1}.
\end{equation}
Following \cite{HSZ} we have introduced the notation

\begin{equation}\label{}
  v = 2\Delta R \geq 1,
\end{equation}
where the inequality follows from the uncertainty principle.

\textbf{4.} In order to calculate the function $G_n(p,p')$ let us note first that
it is the $p,p'$ matrix element of the $n$-th power of the matrix $\rho$.
Therefore, it has the same eigenfunctions and its eigenvalues are $n$-th powers of
the corresponding eigenvalues of matrix $\rho$. Thus

\begin{equation}\label{}
  G_n(p,p') = \sum_{k=0}^\infty \psi_k(p)\psi_k(p')\lambda_k^n =
  \frac{\alpha(1-z)^n}{\sqrt{\pi}}e^{-\frac{1}{2}\alpha^2(p^2 + p'^2)}\Phi(\alpha p, \alpha
  p', \frac{z^n}{2}).
\end{equation}
Substituting the definition of function $\Phi$ and changing variables from $p,p'$
to $K,q$ we find

\begin{equation}\label{}
  G_n(K,q) = \frac{\alpha(1-z)^n}{\sqrt{\pi(1-z^{2n})}}\exp\left[-\alpha^2 K^2
  \frac{1-z^n}{1+z^n} - \frac{1}{4}\alpha^2 q^2\frac{1+z^n}{1-z^n}\right].
\end{equation}
It is convenient to rewrite this formula in the form \cite{HSZ}

\begin{equation}\label{}
  G_n(K,q) = \frac{c_n}{\sqrt{2\pi\Delta^2}}
\exp\left[-\frac{K^2}{2\Delta_n^2} -
  \frac{1}{2}R_n^2q^2\right].
\end{equation}
By comparison with the previous form $\Delta_n = a_n\Delta^2; R^2_n = a_n R^2$
with

\begin{equation}\label{}
a_n = \frac{1}{v}\frac{1+z^n}{1-z^n} = \frac{1}{v}\frac{(v+1)^n + (v-1)^n}{(v+1)^n
- (v-1)^n}
\end{equation}
and

\begin{equation}\label{}
  c_n = \frac{ 2^n\sqrt{v}}{\sqrt{(v+1)^{2n} - (v-1)^{2n}}}.
\end{equation}

\textbf{5.} Let us add some comments.
\begin{itemize}
\item As mentioned, the transition from the one-dimensional case to $d$-dimensions
is very simple. The density matrix (2) gets replaced by a product

\begin{equation}\label{}
  \rho(\vec{p},\vec{p'}) = \prod_{i=1}^d \frac{1}{\sqrt{2\pi\Delta_i^2}}\exp\left[
  - \frac{K_i^2}{2\Delta_i^2} - \frac{1}{2}R_i^2q^2\right].
\end{equation}
Consequently, the integrals $G_m$ from (1) factorize and one obtains for each
$G_m$ a  product of expressions (15), in general with different values of
$\Delta^2$ and $v$ for each factor.
\item For the harmonic oscillator the energy eigenfunctions in the momentum representation
and in the coordinate representation have the same form. Besides the replacement
of $p$ by $x$, the interpretation of the parameter $\alpha$ changes, but this is
irrelevant for the present analysis. Thus, the results given in the present paper
can be used, after some changes in notation, also for the study of Bose-Einstein
correlations in ordinary space.
\item Since the energy eigenvalues of the harmonic oscillator are equidistant, the
corresponding Boltzmann factors form a geometrical progression. Thus, formula (9)
as well as the corresponding formula in the coordinate representation, can be used
to calculate the canonical density matrix for the harmonic oscillator. According
to Landau and Lifszyc \cite{LAL}, this calculation for the diagonal elements of
the density matrix in the coordinate representation has been first performed by F.
Bloch in year 1932.
\end{itemize}

The author thanks Andrzej Bialas for collaboration and stimulating discussions.


\begin{thebibliography} {99}
\bibitem{BAY} G.Baym, Acta Phys. Pol. B29 (1998) 1839.
\bibitem{WHE}U.A. Wiedemann and U. Heinz, Phys. Rep. 319 (1999) 145.
\bibitem{HJA}U. Heinz and B. Jacak, Ann. Rev. Nucl.Part.Sci. 49 (1999) 529.
\bibitem{WEI}R.M. Weiner, Phys. Rep. 327 (2000) 250.
\bibitem{PRA}S. Pratt, Phys. Letters, B301 (1993) 159.
\bibitem{CZI}T. Cs\"org\"o and J. Zim\'anyi, Phys.Rev.Letters 80 (1998) 916.
\bibitem{ZCS}J. Zim\'anyi and T. Cs\"org\"o, Heavy Ion Physics, 9 (1999) 241.
\bibitem{BZ1}A. Bialas and K. Zalewski, Eur.Phys.J. C6 (1999) 349.
\bibitem{BZ2}A. Bialas and K. Zalewski, Phys. Letters 436 (1998) 153.
\bibitem{LED}R. Lednicky et al., Phys. Rev. C61 (2000) 034901.
\bibitem{ZSH}Q.H. Zhang, P. Scotto and U. Heinz, Phys. Rev. C58 (1998) 3757.
\bibitem{HSZ}U. Heinz, P. Scotto and Q.H. Zhang, Ann. Phys. 288 (2001) 325.
\bibitem{WIE}U.A. Wiedemann, Phys. Rev. C57(1998)3324.
\bibitem{RYG}I.S. Gradshtein and I.M. Ryzhyk, Tablitsy integralov, sum, ryadov i
proizwiedienii, 4-th edition Moscow 1962 page 1047.
\bibitem{ABS}M. Abramowitz and I. Stegun, Handbook of mathematical functions,
Dover Publications, New York 1968 page 775.
\bibitem{LAL}L.D. Landau and E.M. Lifszyc, Statisticzeskaja Fizika, Moscow 1964,
page 111.

\end{thebibliography}
\end{document}